# The role of contraction mode in determining exercise tolerance, torque-duration relationship, and neuromuscular fatigue

**Guillaume P Ducrocq, Simon H Al Assad, Nabil Kouzkouz and Thomas J Hureau**

*Mitochondria, Oxidative Stress and Muscular Protection Laboratory (UR 3072), Faculty of Medicine, University of Strasbourg, Strasbourg, France*

*European Centre for Education, Research and Innovation in Exercise Physiology (CEERIPE), Faculty of Sport Sciences, University of Strasbourg, Strasbourg, France*

**Ducrocq GP, Al Assad SH, Kouzkouz N, and Hureau TJ**. *The role of contraction mode in determining exercise tolerance, torque-duration relationship, and neuromuscular fatigue.*
**Purpose**: Critical torque (CT) and work done above it (W′) are key predictors of exercise performance associated with neuromuscular fatigue. The aim of the present study was to understand the role of the metabolic cost of exercise in determining exercise tolerance, CT and W′ and the mechanisms of neuromuscular fatigue.
**Methods**: Twelve subjects performed four knee extension time-trials (6, 8, 10, and 12-minutes) using eccentric, isometric, or concentric contractions (3s-on/2s-off at 90° or 30°/s) to modulate the metabolic cost of exercise. Exercise performance was quantified by total impulse and mean torque. CT and W′ were determined using the linear relationship between total impulse and contraction time. Cardiometabolic, neuromuscular, and ventilatory responses were quantified. Neuromuscular function was evaluated by maximal voluntary contraction, resting potentiated single/doublet electrical stimulations, and superimposed single electrical stimulation to quantify neuromuscular, peripheral, and central fatigue, respectively.
**Results**: Compared to isometric exercise, total impulse (+36±21%; P<0.001), CT (+27±30%; P<0.001), and W′ (+67±99%; P<0.001) were increased during eccentric exercise whereas total impulse (-25±7%; P<0.001), critical torque (-26±15%; P<0.001) and W′ (-18±19%; P<0.001) were reduced in concentric exercise. Conversely, the metabolic response and the degree of peripheral fatigue were reduced during eccentric exercise whereas they were increased during concentric exercise. CT was negatively associated with oxygen consumption gain (R2=0.636; P<0.001) and W′ was negatively associated with rates of neuromuscular and peripheral fatigue indices (R2=[0.252-0.880]; P<0.001).
**Conclusion**: The contraction mode influenced both CT and W′, and consequently exercise tolerance, indicating that the metabolic cost of contraction played a key role.
**Key words**: *Eccentric, isometric and concentric contractions; peripheral and central fatigue; Critical torque and W'; Exercise performance*

**Correspondance to:**
Dr. Guillaume P. Ducrocq g.ducrocq@live.fr

## INTRODUCTION

Determining the physiological mechanisms that limit exercise tolerance or task failure is of particular importance for elite athletes or patients suffering from chronic disease. In the severe-intensity domain, exercise tolerance is well described by the hyperbolic relationship between exercise duration and work rate (1,2). During exercise of a single agonist muscle or muscle group, this relationship is mathematically determined by 1) the critical torque, which represents the asymptote of the hyperbolic relationship and the limit between the heavy and severe intensity domains, and 2) the curvature constant (W′), which represents a fixed amount of work that can be performed above critical torque (1,2).

Critical torque corresponds to a transition phase above which the physiological response to exercise never reaches a steady state (3) and appears to be mainly determined by oxidative metabolism and O2 delivery to the exercising muscles (4,5). During exercise above critical torque, oxidative metabolism is insufficient to supply the exercise ATP demand (5-7). As a consequence, the contribution of non-oxidative metabolisms increases, W′ is progressively used, intramuscular homeostasis is lost and metabolic by-products that are known to impair motor function are continuously accumulated (5-7). As a consequence, motor function is progressively deteriorated during exercise. This phenomenon, identified as neuromuscular fatigue, corresponds to a reversible impairment of subjects' ability to produce force or power and is a potent limiting factor of exercise tolerance (8,9). Neuromuscular fatigue involves the alteration of the muscle's ability to produce force in response to neural inputs (6) and/or the failure of the central nervous system to fully activate the exercising muscle (9). It has been shown that the rate of metabolites accumulation, and thus peripheral fatigue development, depended on the rate of W′ consumption (5,7). Consequently, peripheral fatigue was suggested to be a key limiting factor of W′, particularly during single-limb exercise (4,10,11).

To support this hypothesis, correlative data showed that the degree of peripheral fatigue was associated with W′ between subjects (4,10,11). These results suggest that it is necessary to increase the exercise-induced degree of peripheral fatigue in order to perform a greater W′. However, given the deleterious effect of peripheral fatigue on torque generation capacity (6), this might seem paradoxical. Indeed, one could expect that alleviating peripheral fatigue would enable participants to perform more work, and therefore to increase W′, whereas exacerbating peripheral fatigue would reduce W′. For example, reducing the torque generation capacity with a pre-fatigue reduced W′ during the following exercise (11). Unfortunately, only between-subjects correlations support this hypothesis, and few mechanistic studies, with contrasting results, are available to elucidate the relationship between W′ and peripheral fatigue (4,5,12). Between-subject correlations provide a first step on establishing the physiological mechanisms that link two variables. However, it's not until is conducted a mechanistic manipulation of the variable of interest with a within-subject design, that the validity of the suggested link will be firmly established.

Stronger evidence on the relationship between W′ and peripheral fatigue during single leg exercise might be obtained by modulating the metabolic cost of contraction. This can be achieved experimentally by changing the contraction mode. Indeed, eccentric contractions (i.e. lengthening of the contracting muscles) are characterized by lower metabolic demand than isometric contractions whereas concentric contractions (i.e. shortening of the contracting muscles) are characterized by greater metabolic demand than isometric contractions (13,14). These effects of contraction mode are associated with different exercise tolerance capacity that might be the result of different critical torque and/or W′ (15). However, the effect of contraction mode on critical torque and W′ remains undocumented.

Therefore, the purpose of the present study was to determine the effect of the contraction mode on the parameters of the torque/duration relationship (i.e., critical torque and W′) and elucidate the relationship between W′ and peripheral fatigue. We hypothesized that reducing the metabolic cost of exercise by eccentric contraction improves critical torque and W′ but reduces peripheral fatigue. Conversely, augmenting the metabolic cost of exercise reduces critical torque and W′ but increases peripheral fatigue.

## METHOD
*Participants*
Twelve participants (two women; mean ± SD; age, 23 ± 3 years; height, 176 ± 9 cm; weight, 70 ± 10 kg; body fat, 12 ± 3 %) participated in the present study. The sample size was calculated from the results published by Souron et al. (15) comparing work capacity performed during eccentric and concentric exercises (8.1 x$10^6$ ± 2.7x$10^5$ J vs. 1.7x$10^7$ ± 8.7x$10^6$ J, respectively) with a 5% α error rate and 0.90 β error rate (G*Power 3.1.9.6). All participants gave their written informed consent, and were familiar with intense physical activity which did not involve eccentric contractions specifically. All participants were non-smokers and were non-medicated. The study was approved by the national ethic committee for sport sciences research (IRB00012476-2022-20-03-167) and conducted according to the latest Declaration of Helsinki for human experimentation except for registration in a database.

*Experimental design*
During the preliminary visits, anthropometric measurements were collected. Participants were familiarized with neuromuscular measurements (detailed in "Data collection and analysis" section) before, during and after a single leg exercise of the right quadriceps. After a 3 minutes warm-up at 15% of maximal voluntary contraction (MVC) isometric peak torque (3s on – 2s off), participants performed a 6 minutes maximal single-leg eccentric exercise on an isokinetic dynamometer (Biodex System 3, Biodex Medical Systems, Inc., Shirley, NY, USA) during which they were allowed to freely alter the torque during each contraction. Specifically, participants were asked to perform the greatest amount of work during the imposed time period. Eccentric exercise was chosen for the familiarization trial to elicit a repeated bout effect and limit the muscle damage that could negatively influence exercise performance during the subsequent eccentric experimental visits (16). Range of motion was 90° of amplitude and contraction velocity was set at 30°/s in order to permit a contraction time of 3s. The leg was passively returned to the starting position (just inferior to a fully extended leg) by the dynamometer motor at a velocity of 45°/s in order to permit a 2s recovery time between contraction. Participants were securely fastened by seat belts, arms folded across the chest and were asked to perform as much work as possible. The time elapsed during the exercise was the only information displayed on a large screen chronometer placed directly in front of the participant. Participants were given strong vocal encouragement throughout the exercise.

During the twelve subsequent experimental visits, participants repeated in random order the same procedure but with different exercise durations (i.e. 6 min, 8 min, 10 min, or 12 min) and with different contraction modes (i.e. concentric, isometric, or eccentric). At least 7 days of recovery were allowed after every eccentric visit and at least 2 days of recovery were allowed after every other visit. We chose closed-loop exercises (i.e. self-paced exercise) vs. open-loop exercises (i.e. time to task failure) because exercise performance is less affected by day-to-day variability (17). This procedure has been validated to determine the critical intensity and W′ (18). We chose exercise durations ranging from 6 min to 12 min because it allows an optimal quantification of participants' critical torque and W′ and allows us to study the effect of a very broad spectrum of exercise intensities in the severe-intensity domain (i.e. above the critical power) on neuromuscular fatigue and recovery. During isometric exercises, knee angle was maintained at 90° to coincide with the knee angle set during neuromuscular assessments.

**Data collection and analysis**
*Determination of critical torque and work done above critical torque*
For each exercise, the total torque impulse was calculated by integrating the torque signal during each contraction. The total torque impulse was then plotted against contraction time. The parameters of the torque-duration relationship (i.e. critical torque and W′) were estimated using linear least-square regression with the following equation:

Total torque impulse (contraction time) = W′ + [Critical torque x contraction time]

Where W′ represents the torque impulse done above critical torque.

The averaged standard errors from the model for critical torque and W′ were calculated from the data measured during each time-trial and expressed as the coefficient of variation (CV) The 'total error' of the estimates was calculated as the sum of the standard errors for critical torque and W′.

*Neuromuscular function*
*Contractile function and voluntary activation of the quadriceps.* For the assessment of the contractile function, subjects were seated on the isokinetic dynamometer, arms folded across the chest, with a trunk/thigh angle of 90° and the right knee joint angle at 90° (0° = full extension). The rotation axis of the lever arm was aligned with the lateral epicondyle of the right femur. The lever arm of the dynamometer was fixed to the subject's right ankle, just superior to the malleoli. The settings of the device were carefully recorded and replicated for each subject and visit. The anode, a round self-adhesive electrode, was placed on the femoral triangle, at the stimulation site that resulted in both maximal torque output and maximal amplitude of the compound muscle action potential ($M_{MAX}$) for the vastus lateralis (VL), vastus medialis (VM) and rectus femoris (RF). The cathode, a rectangle self-adhesive electrode, was placed mid-way between the great trochanter and superior iliac crest. The position of these electrodes was marked with indelible ink to ensure a reproducible stimulation site across visits. A constant-current stimulator (DS7AH, Digitimer, Hertfordshire, United-Kingdom) delivered a square wave stimulus (200μs) at a maximum of 400V. To assure maximal spatial recruitment of motor units during the neuromuscular tests, the stimulation intensity was set to 120% of the stimulation intensity eliciting maximal quadriceps twitch and $M_{MAX}$ with increasing stimulus intensities (19). No significant electrical activity of the biceps femoris (BF) was observed during stimulations (120 ± 27 mA; $M_{MAX}$ < 0.5mV).

Neuromuscular assessment of the quadriceps muscle was conducted prior, every 2 minutes during, and immediately after, every single-leg exercise using a standardized set of contractions. During dynamic exercises, neuromuscular assessment was performed by stopping the lever arm at a 90° position for one contraction cycle. In each set, participants performed a 3 s MVC during which a single superimposed twitch ($QT_{single,\ superimposed}$) was delivered at the peak torque of the MVC to determine voluntary activation of the quadriceps (VA, Merton, 1954). Then, potentiated quadriceps twitch evoked by single ($QT_{single}$) and paired [10 Hz ($QT_{10}$) and 100 Hz ($QT_{100}$)] electrical stimulations of the femoral nerve were elicited at 1 s, 3 s, and 5 s after each MVC, respectively. Within-exercise neuromuscular assessments contained only one potentiated quadriceps twitch ($QT_{single}$) to maintain consistent recovery duration (2s) between contractions throughout the entire trial. At baseline, six standardized set of contractions were performed, each one separated by one minute of rest. Following exercise, in order to capture the rapid recovery from fatigue that occurs within the first minutes following exercise termination, the same standardized set of contractions were performed at exactly 0s and 1, 2, 4, 6, 10, and 15 min post-exercise. For all $QT_{10}$, $QT_{100}$, and MVCs, we determined peak torque. For all resting $QT_{single}$, we determined peak torque, contraction time to peak torque, maximal rate of torque development (i.e. maximal value of the first derivative of the torque signal), and half relaxation time (i.e. time to obtain half the decline in maximal torque). The $QT_{10}/QT_{100}$ ($QT_{10:100}$) ratio was calculated, as a decrease in this ratio is commonly interpreted as an index of prolonged low-frequency torque depression (20). Quadriceps VA was calculated according to the following formula: VA (%) = $(1 - QT_{single,superimposed}/QT_{single}) \times 100$. Baseline values for MVC peak torque, maximal rate of torque development, VA, $QT_{10:100}$, $QT_{single}$, $QT_{10}$ and $QT_{100}$ peak torque were calculated by averaging the three highest values from the pre-exercise standardized sets of contractions. Baseline values for twitch contraction time and twitch half-relaxation time were calculated by averaging the three lowest values from the pre-exercise standardized sets of contractions. Percent difference compared to pre-measurements in MVC, VA, $QT_{single}$, $QT_{10}$, $QT_{100}$ and $QT_{10:100}$ were calculated to quantify and characterize the origin of exercise-induced neuromuscular fatigue as follow:

$\Delta$Fatigue(t) (%) = 100 × [Post-Baseline]/Baseline,

where Post represents the value of the index of interest measured at 0, 1, 2, 4, 6, 10 or 15 minutes post-exercise.

The rate of fatigue of neuromuscular indices during exercises was calculated every two minutes as follows: RateOfFatigue (%.min$^{-1}$) = [Xt2 - Xt1]/T with Xt1 and Xt2 corresponding to the exercise-induced reduction in the index (in %) measured at the beginning (t1) and end (t2) of the exercise period (i.e. T = 2min).

To quantify the rate of recovery over time, the time-course of recovery was divided into three periods (12,21): I) from 0 s to 2 min, II) from 2 min to 4 min, and III) from 4 min to 15 min. The recovery rate for each period was calculated as follows: Recovery (%.min$^{-1}$) = 100 × [(Xt1 - Xt2)/XFat]/Trecovery, with Xt1 and Xt2 corresponding to the exercise-induced reduction in the index (in %) measured at the beginning (t1) and end (t2) of the recovery period, XFat corresponding to the highest level of exercise-induced fatigue (i.e. maximal pre- to post-reduction in a given fatigue index; in %), and Trecovery the duration of the recovery period (2, 2 and 11 min for the first, second and third recovery period, respectively).

*Correction for gravity*
The torque signal was corrected to account for the effect of gravity on the lower leg (22). The torque exerted by the lower leg was measured passively at a 30° angle. The 0° (leg fully extended) angle gravity effect on the lower leg was calculated by the following formula:
LegTorqueAt0° = LegTorqueAt30° x 1/cos(30°)
The torque signal was then corrected using the following formula:

CorrectedTorque = Torque + LegTorqueAt0° x cos(KneeAngle)

*Surface electromyography*
Electrical activity of the VL, VM, RF and BF of the right leg was recorded by four pairs of Ag/AgCl surface electrodes (diameter = 10 mm; inter-electrode distance = 20 mm) placed on the muscle belly connected to an EMG system (Octal Bio-Amp - ML138 and PowerLab 16/35, AdInstrument, Bella-Vista, Australia). A reference electrode was placed on the lateral condyle of the right tibia. The skin was shaved, abraded with emery paper and cleaned with alcohol to reduce skin impedance below 3kΩ. The position of the electrodes optimizing $M_{MAX}$ was marked with indelible ink to ensure identical placement at subsequent visits. EMG signals were amplified (gain, 20), filtered (First-order filter, 10 Hz; Fourth order Bessel filter, 500 Hz), and recorded (sampling frequency, 4 kHz) using a commercially available software (Labchart 8, ADInstruments, Bella-Vista, Australia). The rectified EMG signal recorded during each contraction was determined using the onset and offset of the torque signal. The root mean square (RMS) of each burst from the EMG signal was then calculated, normalized to the RMS recorded during pre-exercise MVC ($RMS_{\%MVC}$), and averaged over intervals corresponding to 10% of the total exercise duration. Maximal RMS during each MVC was calculated over the maximal 0.5 s EMG activity.

*Systemic response to exercise*
Pulmonary ventilation and gas exchange indices were measured breath-by-breath at rest and throughout exercises using a stationary automatic ergospirometer (Vyntus MS-CPX, Viasys, San Diego, California, USA). Before each test, gas analyzers were calibrated using a certified gas preparation ($O_2$: 15.9% - $CO_2$: 5%) and an accurate volume of ambient air (2 L) was used to adjust the pneumotachograph. Heart rate (HR) was calculated from a heart rate monitor (V800, Polar Electro, Kempele, Finland). Oxygen uptake ($\dot{V}O_2$), carbon dioxide output ($\dot{V}CO_2$), $\dot{V}CO_2 \cdot \dot{V}O_2^{-1}$, minute ventilation ($\dot{V}_E$), breathing frequency ($fB$), tidal volume ($V_T$) and HR measured during exercise were averaged every 10% of the total exercise duration. Capillary blood samples (5μl) were collected from a fingertip at rest and 3 min post-exercise. Samples were analyzed by an electrochemical method (LactateScout, SensLab GmbH, Germany) immediately after sampling to determine blood lactate concentration ($[La]_b$).

*Near infra-red spectroscopy*
A near infra-red spectroscopy (NIRS; Portamon, Artinis Medical Systems, Netherlands) device was placed on participants' right *vastus lateralis* to assess muscle tissue oxy-, deoxy-, total haemoglobin and myoglobin relative concentrations [heme] and absolute tissue oxygenation ($StO_2$). The NIRS sensors use 3 pairs of LEDs that emit continuous light at 760nm and 850nm. LEDs are placed in a spatially resolved spectroscopy configuration with a source–detector spacing of 30, 35, and 40 mm. The sensor was placed on the *vastus lateralis* proximally to the EMG electrodes locations. Sensor location was marked with indelible ink for reproducible placement between experimental visits and covered with black cloths to avoid signal contamination with ambient light. As recommended by the manufacturer, a differential pathlength factor of 4 was used and data were sampled at 10 Hz. All signals were collected in OxySoft (Artinis Medical Systems, Netherlands).
To account for change in blood volume, oxy[heme] and deoxy[heme] signals were corrected using the following formula (23):

$$\beta(t) = \frac{oxy[heme]}{oxy[heme] + deoxy[heme]}$$

$$oxy[heme]_c(t) = oxy[heme](t) - total[heme(t)] \times (1 - \beta(t))$$

$$deoxy[heme]_c(t) = deoxy[heme](t) - total[heme(t)] \times \beta(t)$$

where $oxy[heme]_c$ and $deoxy[heme]_c$ are the corrected oxygenated and deoxygenated signals, respectively, total[heme] is the blood volume signal from the NIRS device, β is the blood volume correction factor, and t is time. The raw oxy[heme] signal was corrected by subtracting the proportion of the blood volume change attributed to oxy[heme], while the raw deoxy[heme] signal was corrected by subtracting the proportion of blood volume change attributed to deoxy[heme].
At the end of each experimental visit, a blood pressure cuff was placed proximally around the right thigh. The cuff was inflated at suprasystolique pressure (> 300mmHg) to progressively deoxygenate the tissue under the optodes until a plateau in oxy- and deoxy[heme] was observed (~5 minutes). The minimal and maximal oxy[heme] or deoxy[heme] value obtained during this occlusion period and the subsequent hyperemic response were considered as 0% and 100% oxy[heme] or deoxy[heme] (24). Although this procedure neglects the influence of adipose tissue on NIRS data (24), subcutaneous adipose tissue thickness was measured using a skinfold caliper (Harpenden, London, UK) which observed values was divided by 2. All signals were averaged every 10% of total exercises duration.

*Rate of perceived exertion*

To evaluate rate of perceived exertion post-exercise, participants were asked at exercise end to rate on the centiMax scale (CR100) (25) how hard, heavy and strenuous was the exercise. This scale ranged from 0, "nothing at all" to 100, "maximal", with the possibility of rating above 100.

**Statistical analysis**

Data in the Results section are presented as mean ± SD. Because no *exercise duration effect* was found on the main outcome variables (see below for statistical procedures), data collected from different exercise duration were pooled together for figure clarity. Specifically, data collected during the 6, 8, 10, and 12 minutes trials of a given contraction mode were averaged to obtain one data point or one time-series by contraction mode depending on the variable. Normality of every dependent variable and homogeneity of the variance of the distributions (equal variance) were confirmed using the Kolmogorov–Smirnov test and the Levene test, respectively. To protect against the risk of type I error arising from multiple comparisons (26), a multivariate analysis was conducted on our dependent variables recorded during exercise (i.e., mean torque, EMG, and cardiometabolic data) or during post-exercise recovery (i.e., neuromuscular fatigue indices). A significant ($P < 0.001$) *contraction mode x exercise duration x time effect* was found for both the exercise and post-exercise recovery datasets. To determine the effect of contraction mode and exercise duration on time-dependent variables recorded during and following the exercises, three-way ANOVA with repeated measures (*contraction mode x exercise duration x time*) was performed on mean torque, EMG, cardiometabolic data, and neuromuscular fatigue indices. To determine the effect of contraction mode and exercise duration on total impulse, maximal torque output, RPE, [La]b, two-way ANOVA with repeated measures (*contraction mode x exercise duration*) was performed. To determine the effect of contraction mode on critical torque and W′, one-way ANOVA with repeated measures (*contraction mode*) was performed. When a significant difference was found, multiple comparisons analysis was performed using Tukey's HSD test. Effect size was assessed using partial $\eta^2$ ($\eta_p^2$). An $\eta_p^2$ index for effect size was considered as small when $\eta_p^2$ was lower than 0.07, medium when $\eta_p^2$ comprised between 0.07 and 0.20, and large when $\eta_p^2$ was greater than 0.20 (27). Association between peripheral fatigue indices, $\dot{V}O_2$ gain or maximal torque output with W′ or critical torque (i.e., predictors) was tested using linear multilevel models with a random intercept. The regression coefficients (β) were computed with 95% confidence intervals (95% CI), and goodness of the fit from the predictor effect was computed using the $R^2$ coefficient. Statistical analyses were conducted using Statistica v.8.0 (StatSoft, Tulsa, OK, USA), except for the multilevel model analysis, which was conducted using R (v.4.1.2; https://www.R-project.org/) with the packages lme4, lmerTest and modelsummary. Statistical significance was set at $P < 0.05$.

**RESULTS**

*Exercise performance during eccentric, isometric or concentric exercises*

Maximal eccentric contractions elicited a greater peak torque than maximal isometric contractions ($P < 0.001$) whereas maximal concentric contractions elicited a lower peak torque than maximal isometric contractions (Figure 1-A; $P = 0.032$). Total impulse and mean torque were greater during eccentric compared to isometric mode ($P < 0.001$) and during isometric compared to concentric trials ($P < 0.001$, Figure 2-A, Supplemental digital content - Table 1 showing total impulse and mean torque for each exercise contraction mode).

*Critical torque and W′ estimated by eccentric, isometric or concentric exercises*

Eccentric critical torque and W′ were increased by 27 ± 30 % ($P = 0.0097$) and 67 ± 99 % ($P = 0.017$) compared to isometric mode (Figure 1-B/C). Conversely, concentric critical torque and W′ were reduced by -26 ± 15 % ($P = 0.0044$) and -18 ± 19 % ($P = 0.021$) compared to isometric mode (Figure 1-B/C). The changes in critical torque and W′ between contraction modes were significantly associated with the changes in maximal torque (Figure 1-D/E). Critical torque corresponded to 29.0 ± 8.5 %, 30.6 ± 9.7 % and 24.1 ± 6.8 % of the maximal torque developed in eccentric, isometric and concentric mode, respectively. Relative to maximal torque, critical torque was greater during eccentric and isometric compared to concentric mode ($P < 0.001$; $\eta_p^2 = 0.401$, *post-hoc*: $P < 0.01$) with no difference between eccentric and isometric modes ($P = 0.638$). The goodness of fit of the linear Total impulse/Exercise duration model was very good ($r^2 = 0.947$; interquartile range = 0.072). Critical torque was estimated with an error of 3.9 ± 2.3 Nm (CV: 4.6 ± 2.8 %), 3.9 ± 2.8 Nm (CV: 5.7 ± 3.9 %) and 2.7 ± 1.4 Nm (5.3 ± 2.8 %) whereas W′ was estimated with an error of 1211 ± 748 Nm.s (CV: 12.8 ± 11.6 %), 1147 ± 775 Nm.s (CV: 15.6 ± 11.8 %) and 826 ± 407 Nm.s (CV: 14.5 ± 9.0 %) for eccentric, isometric and concentric exercises, respectively. No difference was found between contraction modes for total error of the estimates ($P = 0.722$, $\eta_p^2 = 0.029$).

*Quadriceps muscle activation during eccentric, isometric or concentric exercises*

No pre- to post-exercise difference was found in VL $M_{max}$ (*Time effect*: $P < 0.001$; $\eta_p^2 = 0.652$; *post-hoc*: $P = 0.087$), VM $M_{max}$ (*Time effect*: $P < 0.001$; $\eta_p^2 = 0.608$; *post-hoc*: $P = 0.82$) or RF $M_{max}$ (*Time effect*: $P < 0.001$; $\eta_p^2 = 0.393$; *post-hoc*: $P = 0.27$) suggesting that membrane excitability was not impaired. VL, VM and RF $RMS_{\%MVC}$ were greater during the last 20-30% of exercise in eccentric and concentric compared to isometric mode. In addition, VL, VM and RF $RMS_{\%MVC}$ were greater during the first 10-

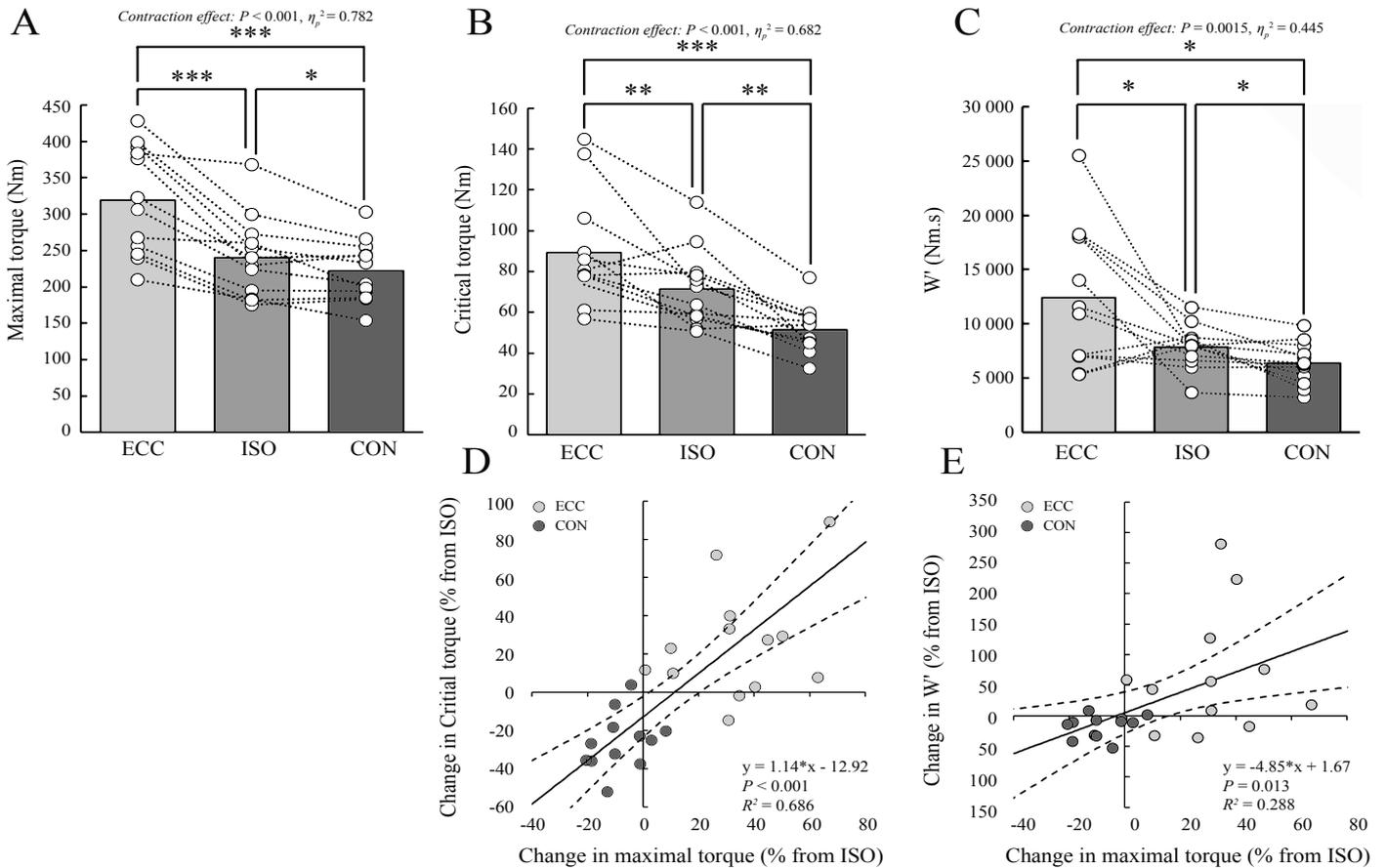

**Figure 1. The effects of the contraction mode on maximal torque (A), critical torque (B) and the work done above critical torque (C)**

Data are presented as group mean (bars) and individuals (white circles and dotted lines) for maximal torque, critical torque and work done above it (W′; upper panels). Results were analysed using one-way repeated measures ANOVA (n = 12).
Associations between the changes in maximal torque, critical torque (D) or W′ (E) with contraction modes (expressed in percent change from ISO) were analysed using linear multilevel models with a random intercept (n = 24; lower panels). The regression line (black line) was computed using the intercept and β coefficient calculated from the models. Dotted lines represent the 95% confidence interval of the regression. ECC, eccentric exercise; CON, concentric exercise; ISO, isometric exercise. *, $P < 0.05$ between the corresponding contraction modes ; **, $P < 0.01$ ; ***, $P < 0.001$

20% of exercise in concentric compared to eccentric mode. For a given torque output, VL, VM and RF muscle activation were greater during concentric compared to isometric mode and during isometric compared to eccentric modes (Figure 2-B/C/D).

*Neuromuscular fatigue induced by eccentric, isometric or concentric exercises*
Except for $QT_{single}$ half-relaxation time that increased, all mechanical indices of neuromuscular fatigue decreased during and following exercises compared to baseline (*time effect*: $P < 0.001$, $\eta_p^2 > 0.71$). All indices remained different than baseline during the 15 minutes of passive recovery (*post-hoc*: $P < 0.01$; Figure 3, Supplemental digital content - Table 2 showing neuromuscular fatigue indices in absolute values).

*Rates of development of neuromuscular fatigue during exercises*

. In isometric and concentric modes, the rates of development of ΔMVC (*time effect*: $P < 0.001$, $\eta_p^2 > 0.77$; *post-hoc*: $P < 0.001$) and $\Delta QT_{single}$ (*contraction mode x time effect*: $P < 0.001$, $\eta_p^2 > 0.42$; *post-hoc*: $P < 0.05$) peak torque reductions were non-linear as they reached their fastest rate of development within the first 2 minutes that followed exercise onset and progressively slowed afterwards. Conversely, in eccentric mode, the rate of development of $\Delta QT_{single}$ peak torque reduction was linear as it did not change throughout exercises
(*post-hoc*: $P > 0.28$). During the first 2 minutes of each exercises, the rate of development of $\Delta QT_{single}$ peak torque reduction were slower during eccentric compared to isometric mode and during isometric compared to concentric mode (*post-hoc*: $P < 0.01$) with no difference afterwards. No difference in the rate of reduction of VA

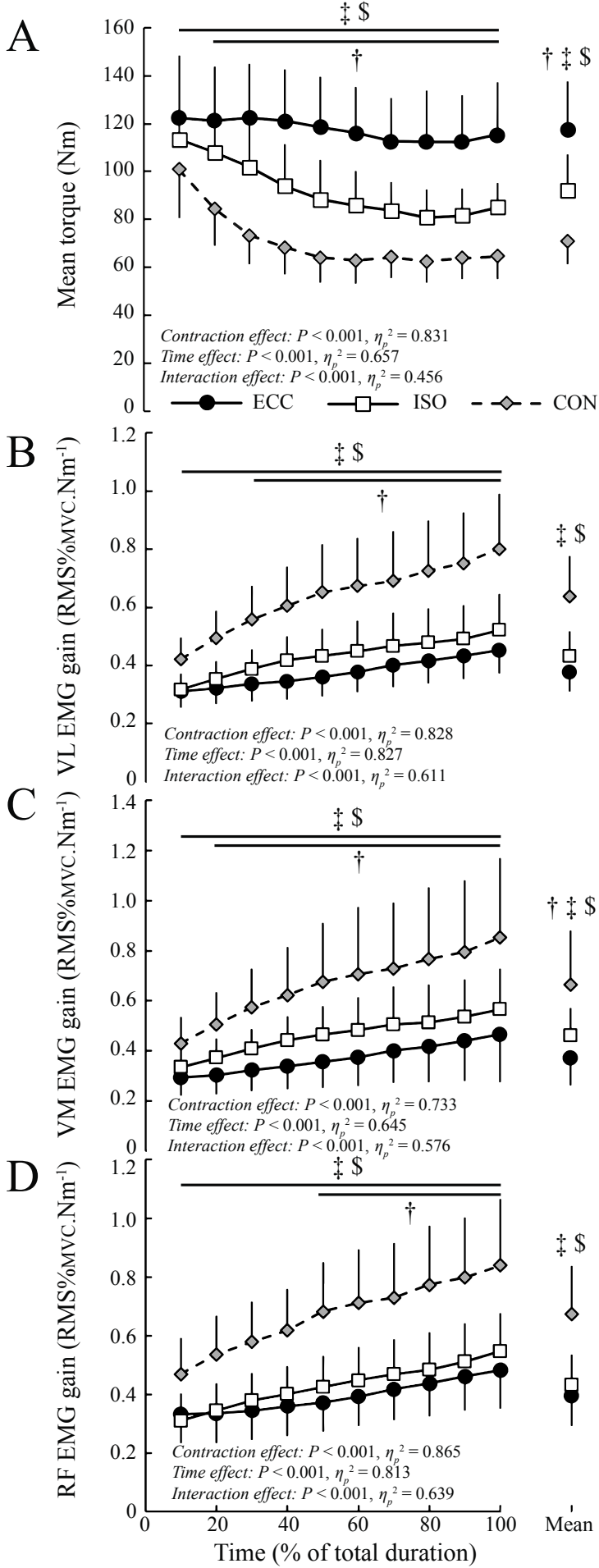

**Figure 2. Effect of the contraction mode on mean torque (A) and quadriceps muscle activation (B-D) during exercise within the severe intensity domain**
Data were analysed using three-way ANOVA (n = 12) and are presented as the mean ± SD.
Muscle activation were calculated with the root mean square (RMS) of the signal recorded from the vastus lateralis (VL), vastus medialis (VM) and rectus femoris (RF). EMG data were normalized to the RMS recorded during pre-exercise MVC and to the mean torque output recorded during the same respective exercise time. MVC, maximal voluntary contraction; ECC, eccentric exercise; ISO, isometric exercise; CON, concentric exercise; EMG, electromyography. †, $P < 0.05$ ECC vs ISO; ‡, $P < 0.05$ ECC vs CON; $, $P < 0.05$ ISO vs CON

was found during exercises (*time effect*: $P > 0.60$, $\eta_p^2 < 0.069$) or between contraction modes (*mode effect*: $P > 0.60$, $\eta_p^2 < 0.069$). The rates of development of ΔMVC and $\Delta QT_{single}$ peak torque reduction were linearly associated with W′ expenditure (Figure 4).

*Neuromuscular fatigue at exercise cessation and during recovery*

A significant *contraction mode x recovery time effect* was found in all neuromuscular indices ($P < 0.008$, $\eta_p^2 > 0.18$) except for VA, VL, VM and RF $M_{max}$ ($P > 0.077$, $\eta_p^2 < 0.11$). Eccentric mode induced a lower ΔMVC, $\Delta QT_{single}$, $\Delta QT_{10}$, $\Delta QT_{100}$ and $\Delta QT_{10:100}$ reduction at exercise cessation compared to isometric and concentric mode, which persisted during the first 2-6 minutes of recovery depending on the index of interest (Figure 3-C/D). Conversely, a greater $\Delta QT_{single}$, $\Delta QT_{10}$, $\Delta QT_{100}$ and $\Delta QT_{10:100}$ reduction was found at exercise cessation for concentric compared to isometric mode, which persisted during the first 2 minutes of recovery.

*Rates of recovery*

A significant *contraction mode x recovery periods effect* was found for all neuromuscular indices ($P < 0.035$, $\eta_p^2 > 0.17$; Supplemental digital content - Table 3 showing rates of recovery from neuromuscular fatigue indices). Except for VA in eccentric mode, all neuromuscular fatigue indices depicted a bi-phasic recovery such as the rates of recovery were faster during the first recovery period (R1) but stopped or became negative during the subsequent recovery periods (R2, R3). Recovery of $\Delta QT_{single}$, $\Delta QT_{10}$, $\Delta QT_{100}$ and $\Delta QT_{10:100}$ were significantly slower during R1 following eccentric compared to isometric or concentric modes whereas no difference was found during R2 or R3. No difference was found between isometric and concentric modes. The rates of recovery of ΔMVC and VA were not different between contraction modes.

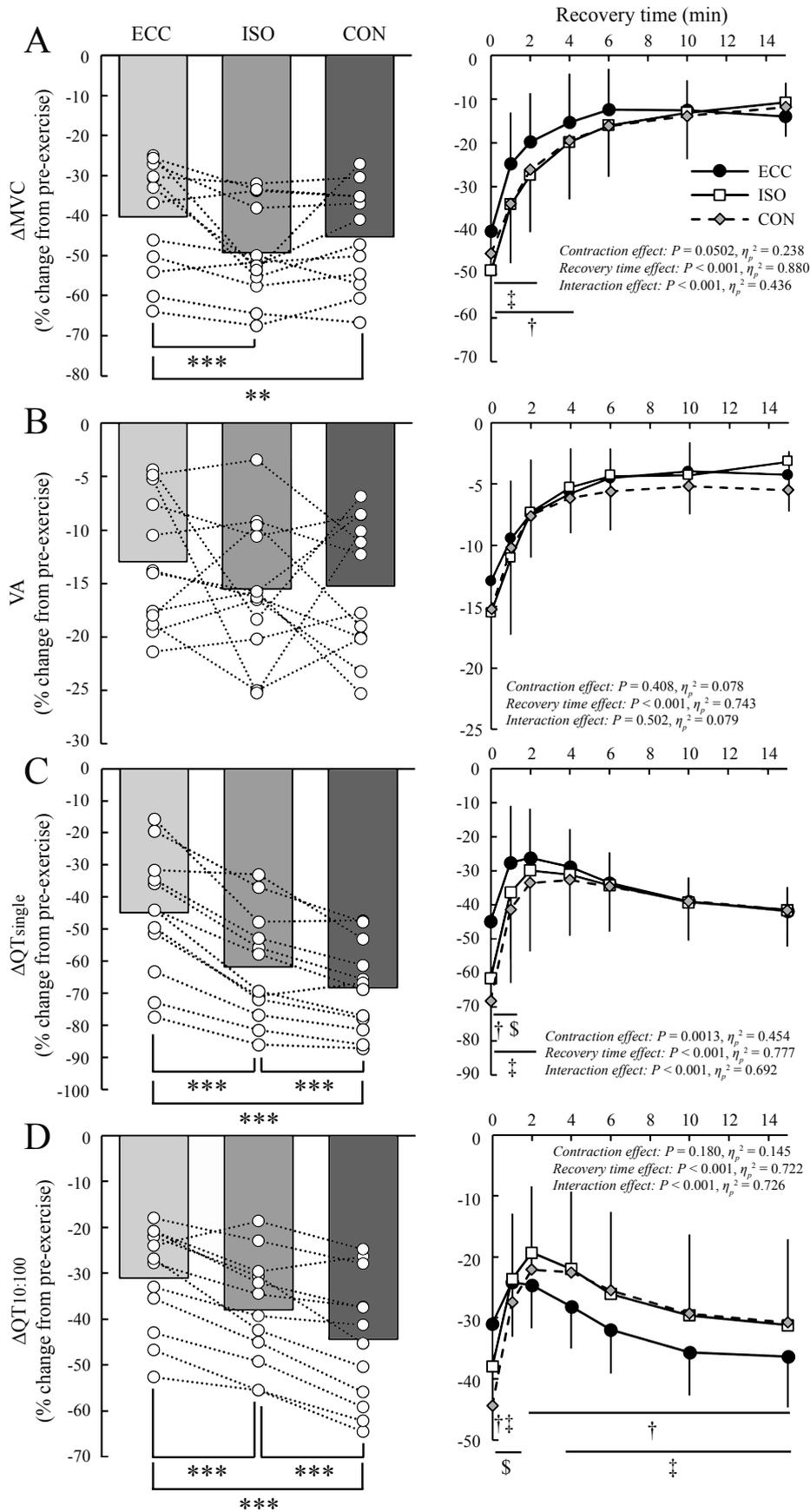

**Figure 3. Effect of the contraction mode on neuromuscular (A), central (B) and peripheral fatigue (C-D) and recovery following exercise within the severe intensity domain**
Group mean (bars) and individual (black circles and dotted lines) data for neuromuscular fatigue indices at exercise termination (left panels) are presented as the percentage change from pre- to 10 s postexercise (except for VA, which remains as a percentage). Group mean ± SD data for recovery from fatigue (right panels) are presented as the percentage change from pre-exercise (except for VA, which remains as a percentage). All data were analysed using three-way ANOVA (n = 12). MVC, maximal voluntary contraction; $QT_{single}$, potentiated twitch peak torque evoked by single electrical stimulation of the femoral nerve; VA, voluntary activation; $QT_{10:100}$, low-frequency fatigue ratio ($QT_{10}/QT_{100}$). **, $P < 0.01$ between the corresponding contraction modes; ***, $P < 0.001$; †, $P < 0.05$ ECC vs ISO; ‡, $P < 0.05$ ECC vs CON; $, $P < 0.05$ ISO vs CON

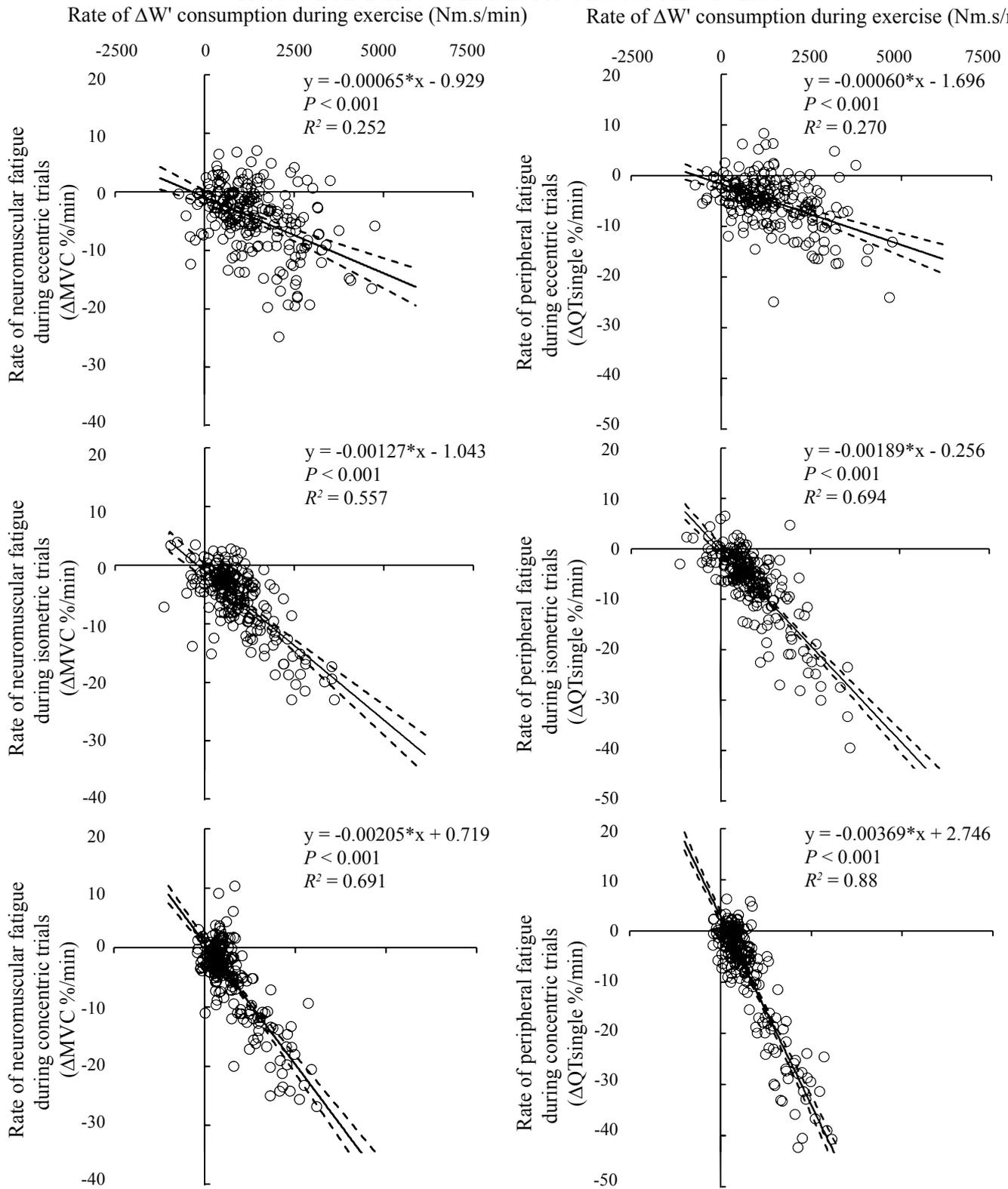

**Figure 4. Association between the rates of development of neuromuscular and peripheral fatigue with the consumption of the work done above critical torque during exercises with different contraction modes within the severe-intensity domain.**
Individual data (open circles) from all exercise (i.e. 6, 8, 10, 12 min exercises) were ploted together. Fatigue data represents the amount of fatigue that has been accumulated per minutes during exercises. This was calculated using the within exercise measurements that were conducted every two minutes (e.g. between minute 2 and minute 4, see method section for further details). W′ data represents the amount of W′ that has been utilized during the same time period. Associations between variables were analysed using linear multilevel models with a random intercept (n = 216). The regression line (black line) was computed using the intercept and β coefficient calculated from the models. Dotted lines represent the 95% confidence interval of the regression. MVC, maximal voluntary contraction; $QT_{single}$, potentiated twitch peak torque evoked by single electrical stimulation of the femoral nerve; W′, work done above critical torque

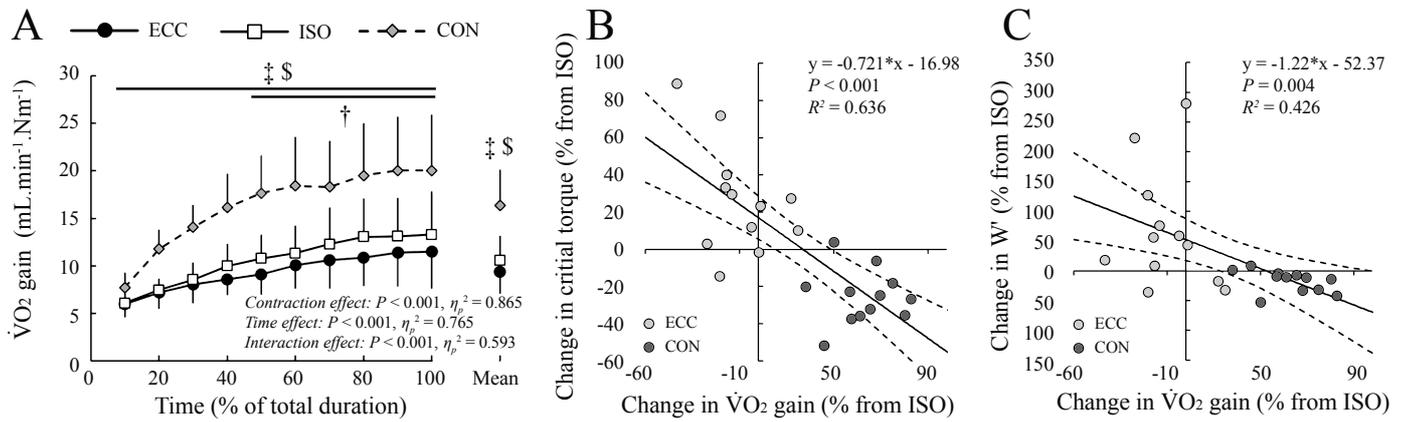

**Figure 5. Effect of the contraction mode on oxygen consumption gain (A) and its relationship with critical torque (B) and the work done above critical torque (C)**

Oxygen consumption ($\dot{V}O_2$) gain was analysed using three-way ANOVA (n = 12) and are presented as the mean ± SD. $\dot{V}O_2$ gain was calculated as oxygen consumption normalized to the mean torque data recorded during the same respective time. Associations between the changes in $\dot{V}O_2$ gain, critical torque or W′ with contraction modes (expressed in percent change from ISO) were analysed using linear multilevel models with a random intercept (n = 24). The regression line (black line) was computed using the intercept and β coefficient calculated from the models. Dotted lines represent the 95% confidence interval of the regression. W′, work done above critical torque; ECC, eccentric exercise; ISO, isometric exercise; CON, concentric exercise; †, $P < 0.05$ ECC vs ISO; ‡, $P < 0.05$ ECC vs CON; $, $P < 0.05$ ISO vs CON

*Systemic responses to exercise*

A significant *contraction mode x time effect* was found for all cardiometabolic and ventilatory indices ($P < 0.01$, $\eta_p^2 > 0.15$) but $V_T$ ($P = 0.489$, $\eta_p^2 = 0.082$). *Post-hoc* analysis revealed that $\dot{V}O_2$ and HR were greater during eccentric and concentric compared to isometric mode (Supplemental digital content - Table 4 showing averaged metabolic and cardiorespiratory response to exercise). When normalized to mean torque output (i.e. $\dot{V}O_2$ gain), $\dot{V}O_2$ gain was 56 ± 17 % greater during concentric mode and -10 ± 18 % lower during eccentric compared to isometric modes (Figure 5-A). The change in $\dot{V}O_2$ gain in eccentric and concentric modes compared to isometric mode was negatively associated with the change in critical torque (Figure 5-B/C). $\dot{V}CO_2$, $\dot{V}_E$, $fR$ and $[La]_b$ were greater during concentric compared to eccentric and isometric exercises. $\dot{V}CO_2/\dot{V}O_2$ was lower during eccentric compared to isometric and concentric modes. No difference was found in RPE between modes of contraction ($P = 0.790$, $\eta_p^2 = 0.045$).

*Muscle oxygenation*

A significant *contraction mode x time effect* was found for all NIRS indices (Figure 6). Specifically, *post-hoc* analysis revealed that oxy[heme] was greater, whereas deoxy[heme] was lower, in eccentric and concentric compared to isometric mode and during the first 10% of eccentric compared to concentric exercises. Total[heme] was lower in eccentric compared to isometric modes and during isometric compared to concentric modes. A greater $StO_2$ was found during the first 10% of eccentric exercise compared to isometric and concentric exercises. No difference was found between isometric and concentric modes ($P > 0.18$).

**DISCUSSION**

Using different contraction modes in order to experimentally manipulate the metabolic cost of contraction, and therefore the development of peripheral fatigue during exercise, we sought to elucidate the determinant of the parameters of the torque/duration relationship and the nature of the relationship between W′ and peripheral fatigue. Consistent with our hypothesis, we found that reducing the metabolic cost of exercise by eccentric contractions was associated with an improved critical torque and W′, and preserved neuromuscular fatigue, compared to isometric exercise. Conversely, increasing the metabolic cost of exercise by concentric contractions was associated with a decreased critical torque and W′, and deteriorated neuromuscular fatigue, compared to isometric exercise. These results indicate that W′ was dictated by the rate at which neuromuscular fatigue accumulated during exercises, the improvement of W′ being associated with slower rates of peripheral fatigue and the deterioration of W′ being associated with faster rates of peripheral fatigue.

**Both critical torque and W′ are sensitive to change in metabolic cost of contraction**

Our findings show for the first time that reducing the metabolic cost of contraction improved exercise performance through greater critical torque and W′. Specifically, critical torque was improved by ~40 % and ~74 % while W′ was improved by ~29 % and ~107 % during exercise in isometric and eccentric modes compared to concentric mode, respectively (Figure 1-B/C). Given the fact that critical torque is determined by the ability of the exercising muscles and the cardiovascular system to deliver and use oxygen to supply ATP demand (4,5), it is not surprising that changing the metabolic cost of contraction have profound effects on critical torque. However, the associated changes in W′ might be less expected. Indeed, these concomitant improvements contrast with previous experiments showing that an improved critical torque/power was compensated by a decrease in W′ and reciprocally (4,5) or that one parameter improved independently of the other (28–30). In the formers, it has been suggested that the mechanisms that were related to the improved critical torque (e.g. increased $O_2$ utilization) had a detrimental effect on W′ (e.g. reduced reliance on non-oxidative metabolic pathways) (4,5). Conversely, our findings demonstrate that reducing metabolic cost of contraction is beneficial for both parameters of the torque/duration relationship and is a key limiting factor of exercise tolerance. Among the different physiological mechanisms that might be responsible for the improved critical torque and W′, we found that, compared to concentric mode, muscle activation was reduced for a given torque output in isometric and eccentric modes (Figure 2-B/C/D). This result suggests that the torque output developed by each motor unit was increased, reducing the number of actin-myosin crossbridge and ATP demand. Greater torque output by the motor units during isometric and eccentric contractions compared to concentric contractions might be the result of the non-ATP dependent engagement of elastic titin proteins by $Ca^{2+}$ released from the sarcoplasmic reticulum (31,32). The predominance of this mechanism is however still vividly debated (31,32) and alternative mechanisms such as non-uniformity of sarcomere length (33) or slower rate of crossbridge cycles during isometric contractions might also play a role (34).

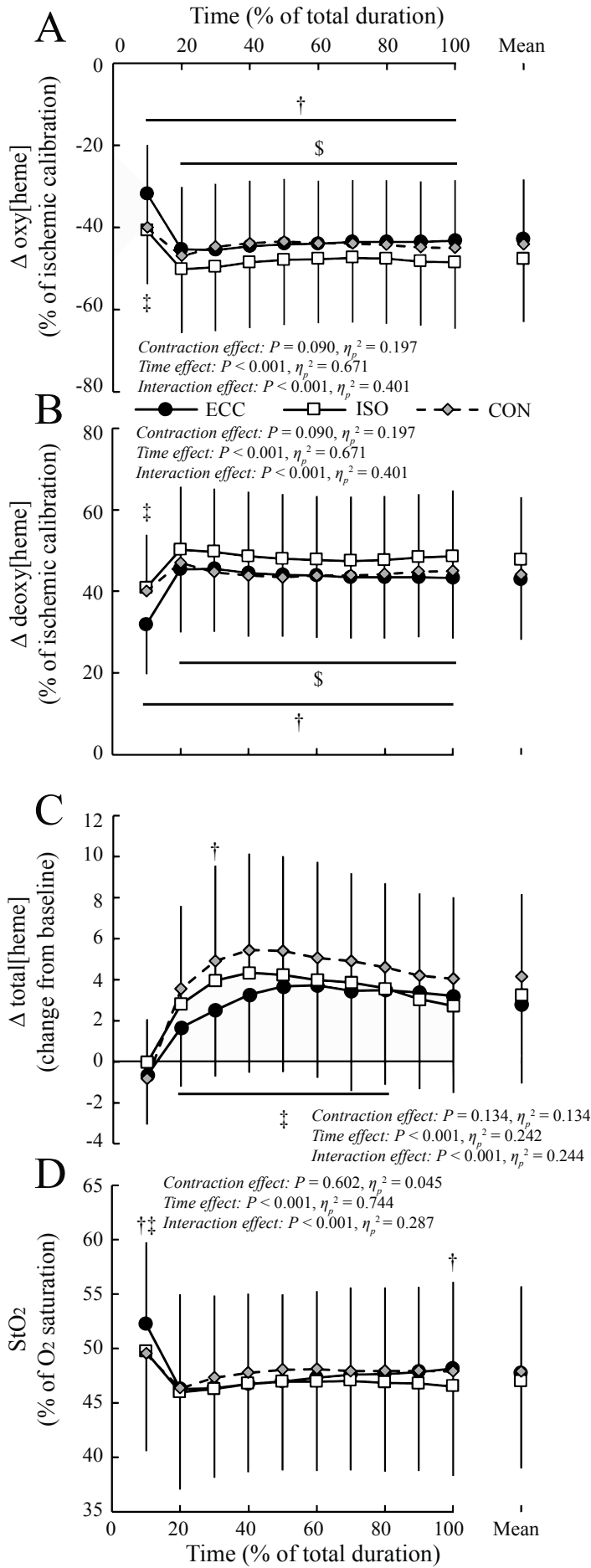

**Figure 6. Effect of the contraction mode on near infrared spectroscopy variables recorded during exercises** Data were analysed using three-way ANOVA (n = 12) and are presented as the mean ± SD. Oxy[heme], concentration of oxygenated heme (hemoglobin and myoglobin, A)); deoxy[heme], concentration of deoxygenated heme (B); total[heme], total amount of heme (oxygenated and deoxygenated, C); $StO_2$, tissue oxygenation index (D); ECC, eccentric exercise; IS, isometric exercise; CON, concentric exercise; †, $P < 0.05$ ECC vs ISO; ‡, $P < 0.05$ ECC vs CON; $, $P < 0.05$ ISO vs CON

**Reducing the oxygen cost of exercise was associated with improved critical torque and W′**

Our data show that greater critical torque was associated with reduced oxidative metabolic demand during exercise in eccentric compared to isometric mode as well as during isometric compared to concentric mode. Indeed, we found that for a given torque output, eccentric contractions consumed ~10% less oxygen whereas concentric contractions consumed ~50% more oxygen than isometric contractions (Figure 5-A). In addition, data extracted from NIRS on the *vastus lateralis* showed that despite greater motor-units activation and torque production in eccentric compared to isometric modes, oxy[heme] and $StO_2$ were greater whereas deoxy[heme] was lower (Figure 6-A/B/C). These results show that $O_2$ utilization from each motor unit recruited was drastically reduced and are in line with previous reports showing lower ATP and/or $O_2$ consumption during eccentric contractions compared to other contractions modes (13,14,35). For example, a previous experiment showed that energy cost of contraction was reduced by ~10% in eccentric, and increased by ~64% in concentric, compared to isometric exercise (13).

Several studies showed that critical torque or power represented a transition phase below which the oxidative metabolic pathway is able to fulfill the ATP demand without sustained contribution of non-oxidative metabolic pathways (3,7). Consequently, reducing the oxidative metabolic demand for a given torque output contributed to shift upward the asymptote of the torque-duration relationship (i.e. critical torque). To support this interpretation, we found that the change in critical torque in eccentric and concentric modes compared to isometric mode was associated with the $\dot{V}O_2$ gain such as reducing oxygen consumption for a given torque output was associated with greater critical torque and inversely (Figure 6). Our interpretation is also supported by previous experiments showing that critical torque/power decreased with arterial occlusion or inspiration of hypoxic gas mixture whereas critical torque increased by inspiration of hyperoxic gas mixture (4,5,29). Finally, previous findings showing that W′ was determined – although to a lower extend than critical torque/power – by oxidative metabolism and intramuscular $O_2$ stores suggest that reducing oxidative metabolic demand also contributed to improve W′ (1,4,5). Indeed, we found that W′ changes between contraction modes was also associated with $\dot{V}O_2$ gain, but the strength of the relationship was weaker than with critical torque (Figure 6). Other mechanisms therefore played a role in changing W′ and our data provide evidence that non-oxidative and neuromuscular mechanisms were predominant.

**Reducing the rate of development of peripheral fatigue was associated with improved W′**

Within the severe-intensity domain, single-leg exercise tolerance is primarily determined by participants' ability and willingness to generate high torque (36). We found that the magnitude of neuromuscular fatigue in eccentric mode was significantly lower than the neuromuscular fatigue induced in isometric mode despite greater torque output. On the opposite, neuromuscular fatigue was not different following concentric and isometric modes despite lower torque output (Figure 3-A). The similar decrease in voluntary activation associated with the smaller degree of peripheral fatigue induced by eccentric compared to isometric and concentric modes suggest that the different degrees of neuromuscular fatigue were the results of a preserved muscle contractile function (Figure 3-C 6). During exercise, the preserved peripheral fatigue consequently favored a greater W′ during exercise in eccentric mode whereas the higher degree of fatigue hampered W′ during concentric compared to isometric mode. As a support for this relationship, we found that the rates of neuromuscular and peripheral fatigue were associated with the rate of W′ consumption such as the greater the amount of W′ was consumed during exercise, the greater the rate of neuromuscular and peripheral fatigue was induced (Figure 4). While this relationship is consistent with previous (4,10,11), but not all (12), investigations, its robustness is questioned by our manipulation of the metabolic cost of contraction through different contraction modes. Indeed, we found that the slope of each relationship decreased from concentric to isometric mode and from isometric to eccentric mode showing that a given W′ consumed did not produce the same degree of neuromuscular and peripheral fatigue contrary to what has been previously suggested (4,10,11).

Given that VL, VM and RF $M_{max}$ remained unchanged, our results suggest that membrane excitability did not contribute to the different potentiated twitch torque reduction (37). However, considering that $\Delta QT_{10:100}$ reduction followed the same pattern as peripheral fatigue indices (Figure 3-D), the different levels of peripheral fatigue were the result of a preserved or deteriorated excitation-contraction coupling function and muscle $Ca^{2+}$ handling/sensitivity (20). The reduced ATP demand associated with eccentric mode compared to other contraction modes has been associated with a spared utilization of phosphocreatine stores and reduced accumulation of intramuscular metabolites (14,38,39). Limiting the accumulation of these potent inhibitors of $Ca^{2+}$ handling and sensitivity likely contributed to preserved contractile muscle function (6). Indeed, we found that $[La]_b$ and $\dot{V}CO_2/\dot{V}O_2$ ratio were reduced during eccentric compared to other contraction modes indirectly suggesting lowered contribution of non-oxidative metabolic pathways and accumulation of intramuscular metabolites. Alternatively, the greater degree of peripheral fatigue during concentric exercises might also be the result of the recruitment and fatigue of a larger pool of fatigue-sensitive motor-units compared to the other contraction modes such as suggested by our EMG data (12,21,40).

**Recovery from neuromuscular fatigue is modulated by contraction mode**

While the effect of contraction mode on the recovery of neuromuscular fatigue has been widely documented for the long-term recovery (15,41,42), the short-term recovery (i.e. the minutes that follow exercise cessation) was scarce. In the present experiment, we found that the rate of recovery of MVC and central fatigue was insensitive to contraction mode whereas the rate of recovery from peripheral fatigue was faster after exercise in isometric and concentric modes compared to eccentric mode (Figure 3-4; Supplemental digital content - Table 3 showing rates of recovery from neuromuscular fatigue indices). We interpret these different rates of recovery as the results of the presumably limited intramuscular metabolite accumulation and spared phosphocreatine stores, as supported by other investigations with direct measures of intramuscular metabolites (43-45). Prolonged low-frequency torque depression has been related to delayed recovery from excitation-contraction coupling failure due to the accumulation of reactive oxygen species (46,47). This mechanism seems unlikely to play a role during eccentric exercises considering the reduced oxidative metabolic activity compared to other contraction modes. Conversely, muscle damage has also been associated with prolonged low-frequency torque depression and it is likely that this mechanism played a role during our experiment despite our attempt to minimize their negative effects (42,48,49).

**Limits and methodological considerations**

We interpreted the changes in critical torque and W′ observed between contraction modes as the result of changing the metabolic cost of contraction. However, modulating the mode of contraction did not only change the metabolic cost of contraction, it also changed contraction velocity and participants' ability to produce maximal torque (15,50,51). Our findings show that maximal torque generating capacity was improved from concentric to isometric mode and from isometric to eccentric mode (Figure 1-A). Our results are consistent with previous findings showing similar results in various muscle groups (48,52) but contrast with data showing that maximal eccentric torque was not different than maximal isometric torque (50,51). This discrepancy might be due to the fact that isometric contractions were not conducted at an optimal angle (i.e. at an optimal quadriceps muscle length) in our experiment leading to an underestimation of maximal isometric torque and possibly critical torque and/or W′. For example, we found that the changes in critical torque and W′ during eccentric and concentric contractions compared to isometric contractions were associated with the changes in maximal torque (Figure 1-D/E). Alternatively, changing contraction velocity might have also contributed *per se* to modulate the torque-duration relationship between contraction modes. Indeed, previous experiments found that reducing pedaling cadence improved critical torque and W' (*i.e.* power normalized by contraction velocity), such as what has been found in isometric compared to concentric mode (53). Overall, these results do not rule out our conclusions but highlight the fact that further data are needed to isolate the role of maximal torque and contraction velocity in determining critical torque and W′.

Although the estimation of the parameters of the power-duration relationship has been validated using whole-body self-paced exercise (18), this method has yet to be specifically validated in single-limb contraction modality. Given the excellent goodness of fit that resulted from the linear model fitting and the low coefficient of variation in the estimation of the parameters, similar to previous studies using constant workload exercise (36,54,55), we are confident that self-paced exercise provided in the present experiment valid measures of critical torque and W′.

**Practical applications**

From a practical point of view, our results provide valuable insights into the limiting factors that dictate exercise performance composed of bouts of mostly eccentric and concentric contractions such as trail running. Indeed, trail running often alternates between uphill, level and downhill runs (56). While uphill running is characterized almost exclusively by concentric contractions, downhill running is characterized almost exclusively by eccentric contractions (57). Participants' critical intensity and ability to perform work above it (i.e. W′) will therefore change dynamically depending on the type of running, providing that exercise intensity is high enough to belong to the severe intensity domain (i.e. short distance trail events or high-intensity interval training sessions).

In addition, our results predict that improving muscle efficiency (i.e. reducing the ATP cost of exercise for a given exercise intensity) would improve the critical intensity (i.e. speed/torque/power) as well as W′. If this hypothesis is verified, the present results will pave the groundwork for future research related to the effect of exercise training on the parameters of the intensity/duration relationship through a change in running/cycling economy.

**CONCLUSION**

Using different contraction modes in order to experimentally manipulate the metabolic cost of contraction, and therefore the development of peripheral fatigue during exercise, we sought to elucidate the determinant of the parameters of the torque/duration relationship and the nature of the relationship between W′ and peripheral fatigue. The greater critical torque found during exercise in eccentric and isometric modes compared to concentric mode was associated with a reduced contribution of oxidative metabolism whereas the greater W′ was associated with a reduced rate of peripheral fatigue. The reduced ATP demand in eccentric mode preserved muscle contractile function through the reduction in intramuscular metabolite accumulation compared to concentric contractions. While we documented a significant correlation between peripheral fatigue and W′, the manipulation of the cost of contraction, using different contraction modes, showed that the slopes of this correlation were different between contraction modes, suggesting that the mechanistic link between peripheral fatigue and W′ is uncertain. These results provide novel insights on the determinant of exercise tolerance. In addition, these findings have important applicative purpose for predicting exercise performance and training prescription of exercises within the severe-intensity domain. Indeed, data collected in one contraction mode cannot necessarily be extrapolated to another exercising condition as their limiting factors might be different.

**COMPETING INTEREST**

The authors have no conflicting interests. The results of the study are presented clearly, honestly, and without fabrication, falsification, or inappropriate data manipulation, and statement that results of the present study do not constitute endorsement by the American College of Sports Medicine.

**ACKNOWLEDGMENTS**

We thank Ms. Lucie Vinot for her valuable assistance in data collection.

**AUTHOR CONTRIBUTIONS**

G.P.D. conceived the study, analysed, interpreted the data and prepared the manuscript. All authors executed the study and edited, revised and approved the final version of the manuscript. All authors agree to be accountable for all aspects of the work in ensuring that questions related to the accuracy or integrity of any part are appropriately investigated and resolved. All persons designated as authors qualify for authorship, and all those who qualify for authorship are listed.

**FUNDING**

This work was supported by the University of Strasbourg Initiative of Excellence (2021-022-10 and ANR-10-IDEX-0002-02).

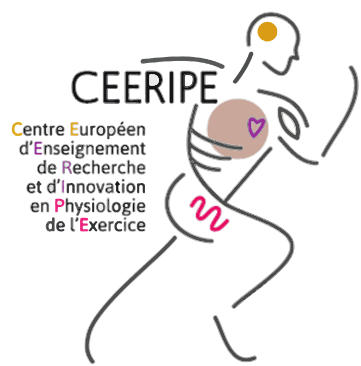

https://ceeripe.unistra.fr

https://f3s.unistra.fr

Supplemental digital content

**Supplemental digitial content - Table 1. Mechanical performance during eccentric, isometric and concentric time-trials**

| Index (units) | Exercise duration | ECC | ISO | CON |
|---|---|---|---|---|
| Total impulse (Nm.S) | 6minTT | 30754 ± 5086 | 22989 ± 4046† | 17308 ± 2257†‡ |
| | 8minTT | 38474 ± 6326* | 29159 ± 5973*† | 21262 ± 3369*†‡ |
| | 10minTT | 45544 ± 8398*# | 33572 ± 6500*#† | 25347 ± 4290*#†‡ |
| | 12minTT | 50807 ± 8263*#$ | 37733 ± 6727*#$† | 28707 ± 4146*#$†‡ |
| Mean torque (Nm) | 6minTT | 127 ± 21 | 100 ± 17† | 78 ± 9†‡ |
| | 8minTT | 121 ± 21 | 97 ± 18† | 73 ± 10†‡ |
| | 10minTT | 117 ± 28* | 90 ± 17*† | 70 ± 13*†‡ |
| | 12minTT | 106 ± 18*# | 83 ± 14*#† | 64 ± 10*#†‡ |

Results are presented as the mean ± SD and were analysed using three-way ANOVA (n = 14). ECC, eccentric exercise; ISO, isometric exercise; CON, concentric exercise; 6minTT, 6 minutes time-trial; 8minTT, 8 minutes time-trial; 10minTT, 10 minutes time-trials; 12minTT, 12 minutes time-trial; *, $P < 0.05$ vs. 6minTT; #, $P < 0.05$ vs 8minTT; $, $P < 0.05$ vs. 10minTT; †, $P < 0.001$ vs. ECC; ‡, $P < 0.001$ vs ISO

**Supplemental digitial content - Table 2. Effects of the exercise mode on indices of neuromuscular function**

| | | ECC | | | ISO | | | CON | | |
|---|---|---|---|---|---|---|---|---|---|---|
| | | | Post exercise | | | Post exercise | | | Post exercise | |
| Index | Units | Pre | 0min | 15min | Pre | 0min | 15min | Pre | 0min | 15min |
| MVC | Nm | 234 ± 55 | 134 ± 27* | 200 ± 49* | 244 ± 57 | 118 ± 15*† | 216 ± 48*† | 232 ± 63 | 121 ± 16*† | 203 ± 48*‡ |
| VA | % | 95 ± 3 | 82 ± 6* | 91 ± 3* | 95 ± 2 | 80 ± 6* | 92 ± 2* | 96 ± 1 | 81 ± 6* | 90 ± 2* |
| $QT_{single}$ | Nm | 55 ± 12 | 29 ± 7* | 31 ± 5* | 55 ± 12 | 20 ± 7*† | 32 ± 4* | 56 ± 12 | 17 ± 5*†‡ | 32 ± 5* |
| $QT_{10}$ | Nm | 90 ± 18 | 42 ± 10* | 42 ± 7* | 90 ± 17 | 32 ± 10*† | 46 ± 8*† | 89 ± 20 | 27 ± 8*†‡ | 46 ± 7*† |
| $QT_{100}$ | Nm | 87 ± 18 | 59 ± 9* | 64 ± 11* | 88 ± 16 | 50 ± 7*† | 66 ± 7* | 87 ± 17 | 47 ± 7*†‡ | 66 ± 9* |
| $QT_{10:100}$ | Ø | 1.04 ± 0.11 | 0.72 ± 0.14 | 0.66 ± 0.12 | 1.03 ± 0.09 | 0.64 ± 0.14† | 0.71 ± 0.15† | 1.02 ± 0.1 | 0.57 ± 0.16†‡ | 0.71 ± 0.15† |
| CT | ms | 87 ± 5 | 64 ± 12* | 62 ± 9* | 87 ± 6 | 71 ± 13*† | 66 ± 9* | 87 ± 7 | 67 ± 18* | 67 ± 12* |
| MRTD | $Nm.s^{-1}$ | 1180 ± 356 | 701 ± 176* | 754 ± 179* | 1211 ± 333 | 447 ± 94*† | 710 ± 123* | 1212 ± 357 | 387 ± 90*† | 755 ± 175* |
| HRT | ms | 71 ± 14 | 72 ± 11 | 50 ± 5* | 71 ± 12 | 91 ± 17*† | 61 ± 7† | 72 ± 13 | 81 ± 18*†‡ | 60 ± 10*† |
| VL $M_{max}$ | mV | 13 ± 3 | 13 ± 3 | 11 ± 3* | 12 ± 3 | 12 ± 3 | 10 ± 3* | 13 ± 3 | 13 ± 3 | 12 ± 3* |
| VM $M_{max}$ | mV | 17 ± 5 | 17 ± 4 | 16 ± 4* | 18 ± 4 | 18 ± 4 | 17 ± 4* | 18 ± 4 | 18 ± 4 | 16 ± 4* |
| RF $M_{max}$ | mV | 6 ± 1 | 5 ± 1 | 5 ± 1 | 6 ± 2 | 6 ± 2 | 6 ± 2 | 6 ± 2 | 6 ± 2 | 6 ± 2 |

Results are presented as the mean ± SD and were analysed using three-way ANOVA (n = 12). No significant difference was found for baseline measurement between conditions. CT, contraction time; HRT, half-relaxation time; Mmax, M-wave maximal amplitude; MRTD, maximal rate of torque development; MVC, maximal voluntary contraction; QTsingle, QT10 and QT100, potentiated twitch peak torque evoked by single, 10 Hz paired and 100 Hz paired electrical stimulation of the femoral nerve, respectively; QT10:100, low-frequency fatigue ratio (QT10/QT100); RF, rectus femoris; VA, voluntary activation; VL, vastus lateralis; VM, vastus medialis; ECC, eccentric exercise; ISO, isometric exercise; CON, concentric exercise. *, $P < 0.05$ vs. Pre; †, $P < 0.05$ vs ECC; ‡, $P < 0.05$ vs ISO

**Supplemental digital content - Table 3. Effects of the exercise mode on recovery of neuromuscular fatigue indices**

| Index (%.min$^{-1}$) | Contraction mode | Recovery periods | | |
|---|---|---|---|---|
| | | R1 (0-2min) | R2 (2-4min) | R3 (4-15min) |
| MVC | ECC | 28.2 ± 14.1 | 6.1 ± 4.7* | -0.4 ± 3.0* |
| | ISO | 22.7 ± 5.5 | 7.8 ± 4.1* | 1.6 ± 1.1* |
| | CON | 22.6 ± 10.9 | 7.5 ± 4.3* | 1.1 ± 1.9* |
| VA | ECC | 13.1 ± 17.5 | 4.7 ± 13 | 0.8 ± 2.2 |
| | ISO | 20.9 ± 12.2 | 5.4 ± 7.4* | 1.4 ± 2.5* |
| | CON | 15.8 ± 14.3 | 2.5 ± 12.8* | 0.2 ± 1.9* |
| $QT_{single}$ | ECC | 19.0 ± 11.4 | -4.2 ± 5.1* | -2.8 ± 1.7 |
| | ISO | 28.6 ± 12.3† | -2.7 ± 7.2* | -1.9 ± 1.3 |
| | CON | 27.1 ± 11.2† | 0.0 ± 3.9* | -1.4 ± 1.5 |
| $QT_{10}$ | ECC | 11.6 ± 7.3 | -3.6 ± 3.5* | -2.0 ± 1.0 |
| | ISO | 22.5 ± 8.9† | -1.9 ± 4.3* | -1.5 ± 0.8 |
| | CON | 23.2 ± 12.4† | 0.1 ± 3.4* | -1.2 ± 0.9 |
| $QT_{100}$ | ECC | 15.9 ± 15.6 | -4.3 ± 6.6* | -0.7 ± 5.2 |
| | ISO | 25.3 ± 10.6† | 0.9 ± 4.2* | -1 ± 1.5 |
| | CON | 25.3 ± 10.8† | 1.4 ± 4.5* | -0.7 ± 1.3 |
| $QT_{10:100}$ | ECC | 7.8 ± 9.3 | -4.8 ± 5.0* | -2. ± 0.8 |
| | ISO | 25.3 ± 9.3† | -4.6 ± 5.2* | -2.1 ± 1.2 |
| | CON | 27.3 ± 12.8† | -0.9 ± 3.4* | -1.7 ± 1.0 |

Results are presented as the mean ± SD and were analysed using three-way ANOVA (n = 12). MVC, maximal voluntary contraction; VA, voluntary activation; $QT_{single}$, $QT_{10}$ and $QT_{100}$, potentiated twitch peak torque evoked by single, 10 Hz paired and 100 Hz paired electrical stimulation of the femoral nerve, respectively; ECC, eccentric exercise; ISO, isometric exercise; CON, concentric exercise. The time course of recovery was divided into the following three periods: (R1) from 15 s to 2 min; (R2) from 2 to 4 min; and (R3) from 4 to 15 min. *, $P < 0.05$ vs. R1; †, $P < 0.05$ vs. ECC;

**Supplemental digitial content - Table 4. Effects of the exercise mode on cardiometabolic, respiratory and perceptual responses**

|  | Rest | ECC | ISO | CON |
|---|---|---|---|---|
| $\dot{V}O_2$ (mL.min$^{-1}$) | 450 ± 71 | 1076 ± 333 | 936 ± 275† | 1101 ± 285‡ |
| $\dot{V}CO_2$ (mL.min$^{-1}$) | 387 ± 67 | 1066 ± 362 | 983 ± 288 | 1169 ± 336†‡ |
| $\dot{V}CO_2/\dot{V}O_2$ | 0.86 ± 0.04 | 0.98 ± 0.08 | 1.05 ± 0.10† | 1.05 ± 0.08† |
| $\dot{V}_E$ (L.min$^{-1}$) | 15 ± 2 | 36 ± 13 | 33 ± 10 | 40 ± 13†‡ |
| $fR$ (breaths.min$^{-1}$) | 21 ± 2 | 21 ± 7 | 20 ± 7 | 23 ± 8†‡ |
| $V_T$ (L) | 1.5 ± 0.3 | 3.8 ± 1.2 | 3.6 ± 0.98 | 3.7 ± 1.0 |
| HR (beats.min$^{-1}$) | 70 ± 10 | 120 ± 18 | 115 ± 19† | 119 ± 17‡ |
| [La]$_b$ (mmol.L$^{-1}$) | 1.1 ± 0.1 | 3.4 ± 2.1 | 3.8 ± 1.8 | 4.8 ± 2.5†‡ |
| RPE (A.U) | N.A. | 92 ± 6 | 92 ± 8 | 93 ± 7 |

Results are presented as the mean ± SD and were analysed using three-way ANOVA (n = 12). $\dot{V}O_2$, oxygen consumption; $\dot{V}CO_2$, $CO_2$ output; $\dot{V}_E$, minute ventilation; $fR$, breathing frequency; $V_T$, tidal volume; HR, heart rate; [La]$_b$, capillary blood lactate concentration; RPE, Rate of perceived exertion; †, $P < 0.05$ vs ECC; ‡, $P < 0.05$ vs ISO